# Mission Opportunities for Human Exploration of Nearby Planetary Bodies

Cyrus Foster† and Matthew Daniels‡
*Stanford University, Stanford, California 94305*

**We characterize mission profiles for human expeditions to near-Earth asteroids, Venus, and Mars. Near-Earth objects (NEOs) are the closest destinations beyond cis-lunar space and present a compelling target with capabilities already under development by NASA and its partners. We present manned NEO mission options that would require between 90 days and one year. We next consider planetary flyby missions for Venus along the lines of plans that were first drafted during the Apollo program for human exploration of Venus. We also characterize a Mars flyby, and a double-flyby variant that would include close passes to both Venus and Mars. Finally, we consider orbital missions to Venus and Mars with capability for rendezvous with Phobos or Deimos. This would be a truly new class of mission for astronauts and could serve as a precursor to a human landing on Mars. We present launch opportunities, transit time, requisite ΔV, and approximate radiation environment parameters for each mission class. We find that ΔV requirements for each class of mission match near-term chemical propulsion system capabilities.**

## Nomenclature

| | | |
|---|---|---|
| *AU* | = | Astronomical Unit, or Earth-Sun Distance |
| *C3* | = | Characteristic energy ($km^2/s^2$) |
| *ΔV* | = | Delta-V, change in velocity (km/s) |
| *LEO* | = | Low Earth orbit |
| *NEA* | = | Near Earth Asteroid |
| *NEO* | = | Near Earth object |
| *PHA* | = | Potentially hazardous asteroid |
| *SPK* | = | Spacecraft and Planet Kernel |

†M.S. Student, Department of Aeronautics & Astronautics, Stanford University. Education Associate, NASA Ames Research Center. AIAA Student Member, Cyrus.Foster@gmail.com.
‡Ph.D. Student, School of Engineering, Stanford University. Aerospace Engineer, Mission Design Center, NASA Ames Research Center. AIAA Student Member. Matthew.P.Daniels@nasa.gov.

## I. Introduction

Destinations for human exploration beyond Earth orbit can serve as stepping stones for operations and technology development in NASA's human spaceflight program toward the eventual human exploration of Mars. A human presence beyond the Earth-Moon system would generate scientific returns not afforded by experiments in LEO. Likewise, it would bring educational benefits more aligned with the Apollo program: educating young people by putting them to work on a grand challenge, and inspiring students through the accomplishments of a technology vanguard in space exploration. This is a timely issue for consideration as a more precise definition of the Augustine Commission's flexible path concept, and political debate about the future direction of human spaceflight policy, take on programmatic urgency for NASA and the American space industry.

We characterize mission profiles for human expeditions to near-Earth asteroids, Venus, and Mars. NEOs are the closest destinations beyond cis-lunar space and present a compelling target with capabilities already under development by NASA and its partners. The composition and origin of near-Earth asteroids are still not well understood, and are a subject of considerable scientific interest; likewise, the low-probability but high-consequence catastrophe of an asteroid impact on Earth poses a substantial risk, as described by the 2010 National Academies report *Defending Planet Earth: Near-Earth Object Surveys and Hazard Mitigation Strategies*[1]. Asteroids offer a compelling human exploration target for scientific and planetary risk reasons; these missions also present an opportunity to take the first technological and operational steps for human expeditions into the solar system.

We consider planetary flyby missions for Venus and Mars, along with a double-flyby variant mission that would include close passes to both planets. We find that a Venus flyby mission and certain double-flyby opportunities require a duration shorter than the current human record for continuous time in space of 438 days set in the 1990s on Mir[2]. Finally, we consider orbital missions to Venus and Mars with capability for rendezvous with Phobos or Deimos.

We present mission opportunities during the 2015-2025 decade in the appendix and include summarizing mission parameters such as the launch frequency, duration and propulsion requirements.

## II. Methodology

This section discusses the algorithmic approach to trajectory generation for NEO and planetary missions. For mission opportunities skip to *Sections III-V*.

*NEO Missions*

We generate flyby and rendezvous missions to NEOs by sequentially solving Lambert's problem[3] with departure dates across the 2015-2025 decade and mission durations of 90, 180 and 365 days. We utilize these mission durations to provide a basis of comparison for NEO flyby and rendezvous mission requirements; in practice, a mission can be optimized to specific parameters of concern.

A MATLAB program first pulls orbital parameters for each of the 7054 known NEOs in the JPL *Small Body Database*[4] as of July 1, 2010. The program generates state vectors using two-body propagation for each object in 1.25-day time steps across the time period of 2015-2025. Similarly, Earth's orbital parameters are stored for each 1.25-day time period in this domain. For each departure date, the MATLAB program then steps through 72 time steps (for 90-day mission), 144 time steps (180-day mission), or 292 time steps (365-day mission), solving Lambert's problem for the Earth-NEO pair to generate mission $\Delta V$ values at each time step. Trajectories are stored subject to meeting constraints on $\Delta V$ as follows: post-escape $\Delta V$ must be less than 4 km/s for a flyby mission or 7 km/s for a rendezvous mission; Earth departure C3 must be less than 15 $km^2/s^2$ ($\Delta V$ of 3.9 km/s from a 200 km LEO); and reentry inertial speed must be less than 13.4 km/s. The minimum-$\Delta V$ opportunity is stored for the corresponding NEO and mission duration-class (e.g., a flyby of asteroid 7482 on 19-Jan-2022 as a 90-day mission). This brute force search is run across each NEO and mission duration, producing the statistics for all 90-, 180-, and 365-day missions to 7054 known NEOs during the 2015-2025 time span.

For the rendezvous missions, we picked the 200 lowest-$\Delta V$ candidates and re-ran the sequential Lambert solver using SPK ephemerides generated by the JPL *HORIZONS* system[5] producing the results in table 2 and the appendix. For flyby missions, the detailed results presented in appendix A were also evaluated with SPK data, while table 1 shows the results of a broader search using the less exact osculating elements.

*Planetary Missions*

We generate the planetary missions using the same method as for the NEO missions but without round-trip duration constraints. A full-factorial search is performed using the DE421 planetary ephemeris[6] and 4-day time steps. We calculate flyby-ΔV and altitude from the inbound and outbound excess hyperbolic velocities and required turn angle, assuming the ΔV impulse is made at periapsis[7]. A particular flyby scenario is ignored if the required periapsis is below the relevant altitude presented in table 3.

From the Lambert solver data, we generate contour plots of constant total mission ΔV as a function of Earth departure date and mission duration. (These plots are colloquially known as 'pork chop' plots, where each 'pork chop' represents a launch opportunity, such as a Venus flyby mission every 1.6-year synodic period). For each launch opportunity a Pareto front of total-ΔV versus mission duration is computed. The mission opportunity results presented in this paper are taken from the Pareto front by minimizing a linear combination of ΔV and duration (ΔT), such that the trajectory minimizes the sum $\alpha \Delta V + \Delta T$. This approach selects missions close to minimum energy but is calibrated to seek opportunities with shorter duration at small ΔV costs. This is illustrated in figure 1. The constant α is made to be a function of the energy requirements of the mission, such that: $\alpha = \gamma * \min(\Delta V)$. For this paper γ is set by inspection to equal 0.04. This α allows the relative weight of ΔV and ΔT to vary with mission class; for more energy-intensive missions (higher minimum ΔV), α weights more favorably missions which don't increase ΔV further in search of time-savings. This process of mission selection from the Pareto front can be calibrated differently, or replaced by human judgment, depending on user needs.

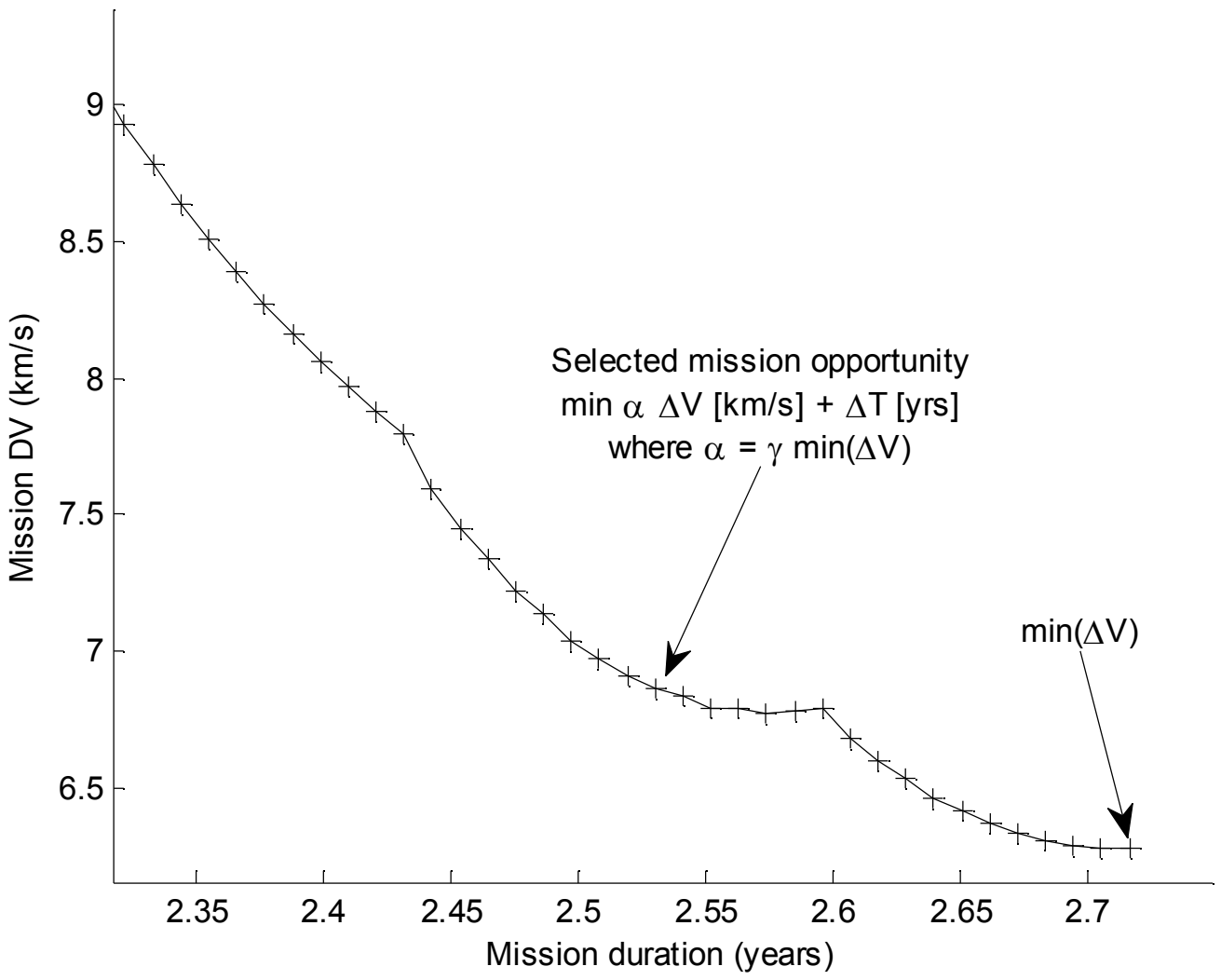

**Figure 1**. Pareto front for the August 2020 Mars long-stay mission indicating the formula used to select opportunities. For this paper, γ is set to be 0.04.

*Phobos and Deimos*

An analytical 2-D study was conducted to establish ΔV and time requirements for visiting Phobos and Deimos from a Mars staging orbit. For such transfers, one has two options: thrusting at apoapsis to raise periapsis to intersect the moon of interest; or alternatively, to thrust at periapsis to lower apoapsis to the moon's orbit. These routes each require two impulses and have different ΔV requirements. For transfers between the staging orbit and Martian moons, we find the lowest-ΔV strategy calls for thrusting at the staging orbit apoapsis, raising periapsis to intercept either moon. If aerobrake capability through the Martian atmosphere is available, however, the ΔV requirement for reaching Phobos is minimized by lowering the apoapsis via aerobrake (this is not the case for Deimos).

When performing a Mars escape impulse, we find the lowest-ΔV option is to return to the staging orbit and make an escape burn at periapsis, rather than performing the escape maneuver directly from either of the Martian moons' orbits.

## III. Near Earth Object Missions

We consider a NEO population composed of 7054 near-Earth asteroids (NEAs) and 85 near-Earth comets known by the NASA Near Earth Object Program[8] as of July 1st 2010. Of this NEA population, 814 are larger than 1 km in diameter. The National Research Council *Committee to Review Near-Earth-Object Surveys and Hazard Mitigation Strategies* estimates that nearly 85% of objects 1 km in diameter or larger in the near-Earth environment are now known, and that NEOs larger than 1 km in diameter (not including long-period comets) should number about 940[1].

The discovery rate of sub-kilometer NEOs, in contrast, is dramatically increasing. These bodies are also uniquely compelling due to their high probability of collision with Earth. Near-Earth asteroid 99942 Apophis, with a diameter of approximately 300 m, will pass within 36,000-40,000 km from the Earth's center on Friday, April 13, 2029 at about 21:45 UTC – a distance comparable to the geostationary orbit where communications satellites are frequently placed. During its closest approach, Apophis will be a 3rd-magnitude object, making it visible to the unaided eye from Asia, Africa, and Europe[9]. It is now established with some confidence that Apophis will not strike Earth in 2029; impact in 2036, however, has not been ruled out due to uncertainties about how Apophis' composition and interaction with sunlight affect its orbit[1]. This motivates a capability to visit and study NEAs for planetary defense reasons.

### A. NEO Flyby Missions

We present notional mission opportunities with round-trip durations of 90, 180 and 365 days. In practice, a mission would be optimized to specific parameters of concern, and the total ΔV and duration would change. One can usually reduce the mission duration at the cost of increased ΔV, or conversely reduce the required ΔV if a longer mission duration is acceptable. Table 1 indicates the total number of NEO targets accessible via 90, 180, and 365 day missions, binned according to four post-escape ΔV ranges. Post-escape ΔV indicates the impulsive maneuver capability required after the flyby for Earth return.

There is generally an opportunity for a low-ΔV flyby mission when a NEO makes a close approach to the Earth. A spacecraft will escape the Earth-Moon system into a heliocentric orbit adjacent to Earth's to intercept the NEO during its close-approach, and then fire an impulse to place it on a return trajectory. A 90-day mission with a ΔV less than 1 km/s will typically reach no more than 10 lunar distances from Earth, limiting possible flyby targets to potentially-hazardous asteroids (PHAs). Longer missions can reach more distant targets under the same ΔV budget; for example a 365-day mission requiring 1 km/s could reach NEO targets as much as 0.3 AU distant.

The advantage of a flyby over a rendezvous is that any NEO making a close-approach is a potential candidate regardless of its orbital parameters. Figure 2 shows the heliocentric trajectory of a 90-day flyby mission to Asteroid 7482 (1994 PC1), estimated to be 1-2 km in size[10]. Although this asteroid has an orbital inclination of 33.5° making it inaccessible from a practical rendezvous mission, its intersection with the ecliptic plane in January 2022 at close proximity to the Earth creates the opportunity for a flyby mission. The diversity of NEOs accessible via a flyby during a 10-year span is evident by the fact that almost 50% of known NEOs can be reached on a 365-day mission with a post-escape ΔV less than 1 km/s. This includes NEOs over 10 km in size; those that are highly inclined or highly eccentric; targets of various spectral types; and comets. Appendix A provides some examples of interesting flyby targets.

**Table 1.** Total number of flyby mission opportunities to NEOs during the 2015-2025 decade with an Earth-escape C3 less than 15 $km^2/s^2$ ($\Delta V$ of 3.9 km/s from a 200 km LEO) and aerocapture inertial speed for Earth return less than 13.4 km/s. If a NEO can be visited on more than one occasion for a given mission duration, the lowest $\Delta V$ opportunity per mission duration is selected.

**Total Number of Flyby Mission Opportunities**
**by Post-Earth-Escape ΔV and Mission Duration for 2015-2020**

| **Post-escape ΔV** | **90-day mission** | **180-day mission** | **365-day mission** |
|---|---|---|---|
| $\Delta V < 1$ km/s | 26 | 411 | 3144 |
| 1 km/s $< \Delta V <$ 2 km/s | 81 | 438 | 301 |
| 2 km/s $< \Delta V <$ 3 km/s | 112 | 407 | 268 |
| 3 km/s $< \Delta V <$ 4 km/s | 90 | 280 | 228 |
| **Total Opportunities** | **309** | **1536** | **3941** |

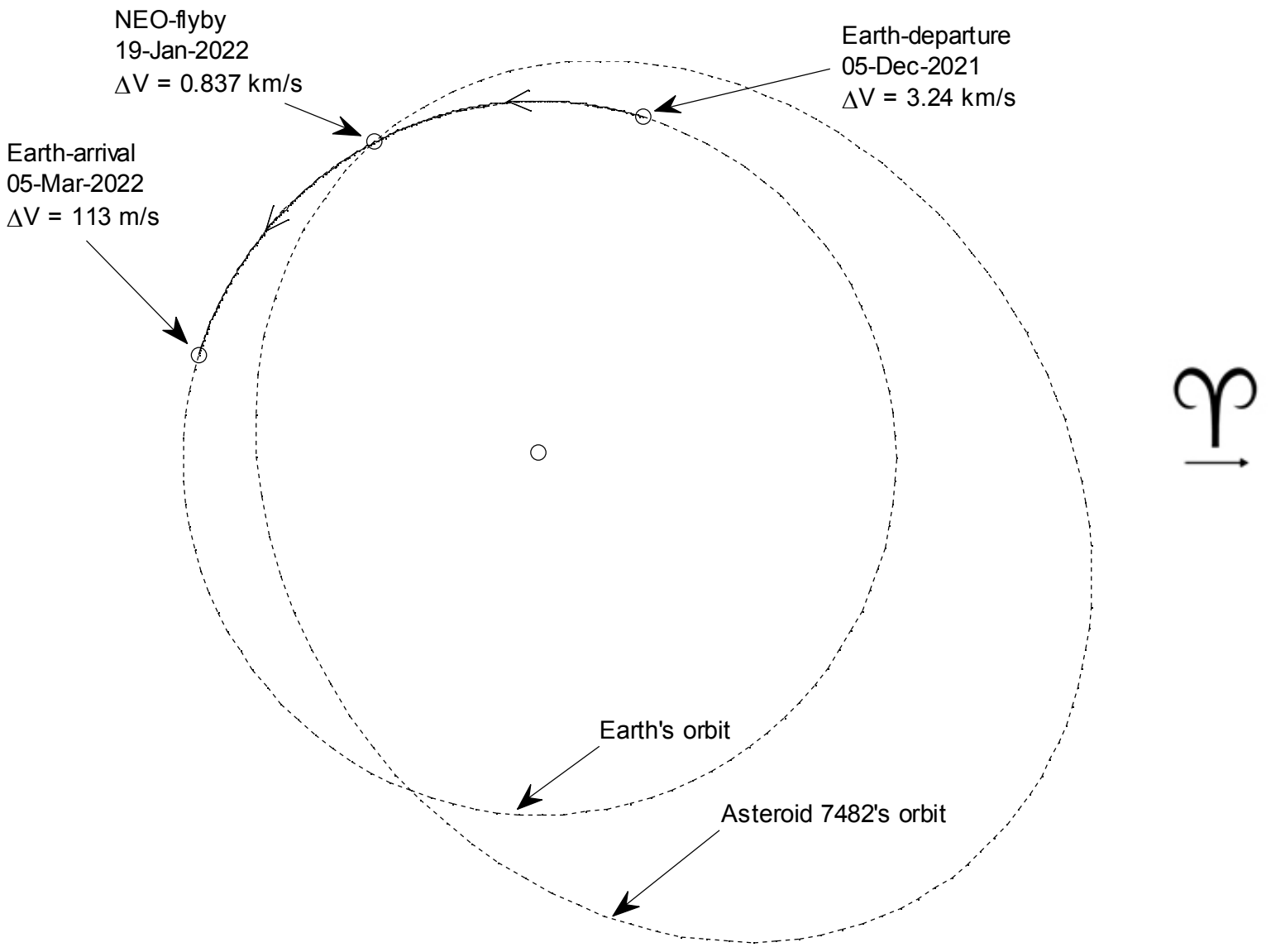

**Figure 2**. Trajectory of a 90-day flyby mission to Asteroid 7482 (1994 PC1), indicating event dates and required ΔV. The spacecraft's trajectory closely follows the Earth, deviating only enough to intercept the asteroid as it makes a close approach to the Earth. Departure ΔV is from a 200 km LEO; Earth-arrival ΔV is the aerocapture load for insertion into an Earth staging orbit (see *Section V. A.;* to calculate inertial speed, add 10.91 km/s to the ΔV aerocapture load).

## B. NEO Rendezvous Missions

Table 2 indicates the total number of NEO targets accessible during the 2015-2025 decade under three mission regimes: 90-day missions with 5-day stay, 180-day missions with 10-day stay, and 365-day missions with 20-day stay. As with flybys, rendezvous missions can be optimized on a case-by-case basis, with trades in stay time, round-trip duration and ΔV. These opportunities are sorted by post-Earth-escape ΔV, the impulsive maneuver capability required for arrival at and departure from the NEO. We find a total of eight missions of 90-day duration, with a lowest post-escape ΔV of 2.56 km/s; 28 missions of 180-day duration, with a lowest post-escape ΔV of 1.97 km/s; and 107 of 365-day duration, with a lowest post-escape ΔV of 1.01 km/s. Longer missions of 1.5- or 2-year duration can also be designed with lower ΔV requirements and the ability to reach a wider range of NEO targets.

The lowest ΔV targets are small NEAs with similar orbital parameters to the Earth, such as asteroid 1991 VG in figure 3. The largest 90-day target is 2001 QJ142 with a diameter on the order of 100 m, while most candidates in that mission class are only approximately 10 m in size. Most of these objects have only recently been discovered: five out of the eight 90-day targets listed in table 2 were discovered since the latest publication on human NEO rendezvous missions in 2007[11]. One can speculate that many more small-NEA rendezvous targets will be discovered in the years ahead (at the current rate of discovery, the number of known NEOs will at least double in the next decade).

We provide additional details on a list of favorable NEO rendezvous targets in Appendices A and B.

**Table 2.** Total number of rendezvous mission opportunities to NEOs during the 2015-2025 decade with an Earth-escape C3 less than 15 $km^2/s^2$ and aerocapture inertial speed for Earth return less than 13.4 km/s. If a NEO can be visited on more than one occasion for a given mission duration, the lowest ΔV opportunity per mission duration is selected.

**Total Number of Rendezvous Mission Opportunities**
**by Post-Earth-Escape ΔV and Mission Duration for 2015-2020**

| **Post-escape ΔV** | **90-day mission with 5-day stay** | **180-day mission with 10-day stay** | **365-day mission with 20-day stay** |
|---|---|---|---|
| ΔV < 1 km/s | 0 | 0 | 0 |
| 1 km/s < ΔV < 2 km/s | 0 | 1 | 5 |
| 2 km/s < ΔV < 3 km/s | 1 | 2 | 4 |
| 3 km/s < ΔV < 4 km/s | 0 | 2 | 15 |
| 4 km/s < ΔV < 5 km/s | 1 | 7 | 20 |
| 5 km/s < ΔV < 6 km/s | 3 | 9 | 30 |
| 6 km/s < ΔV < 7 km/s | 3 | 7 | 33 |
| **Total Opportunities** | **8** | **28** | **107** |

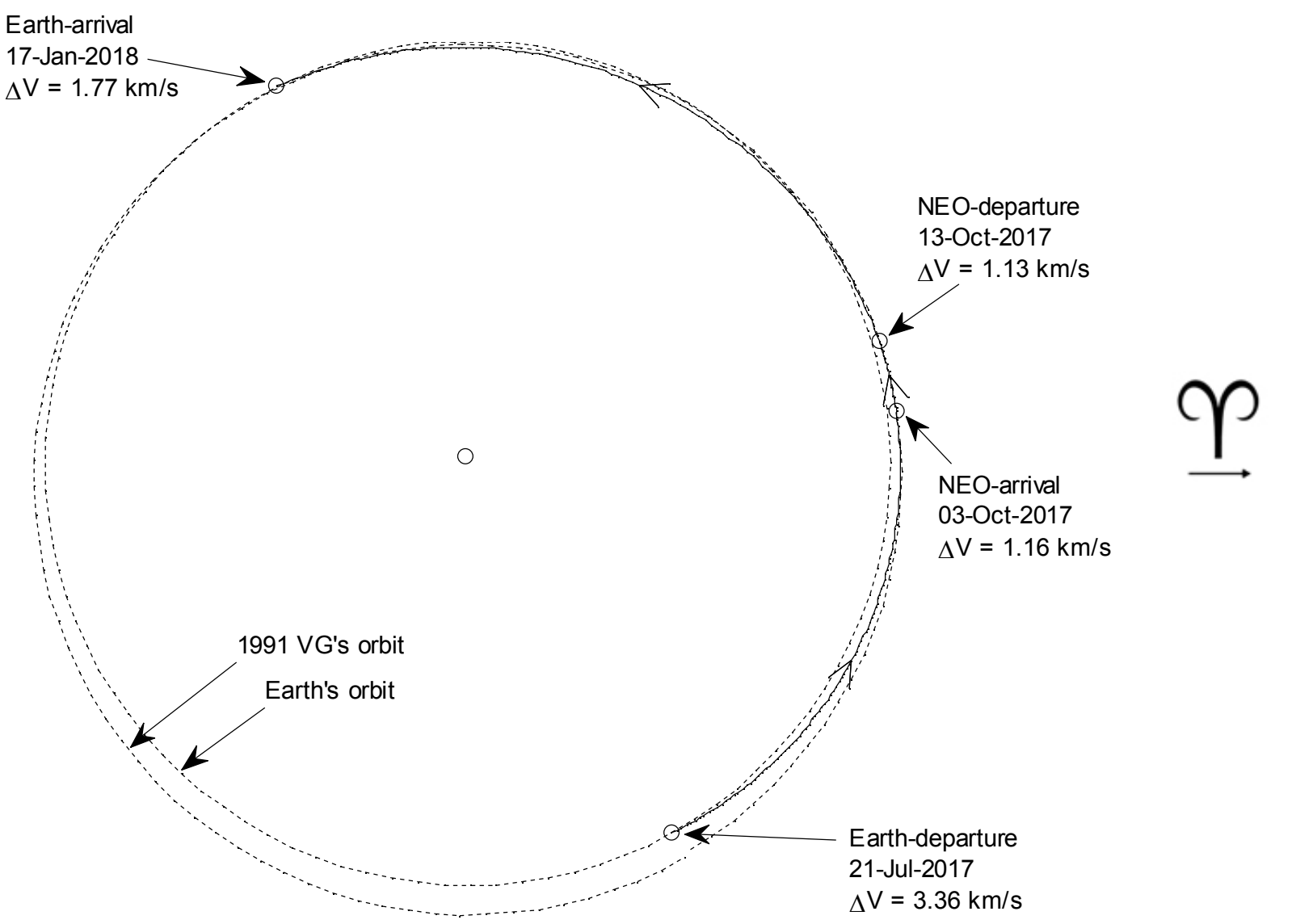

**Figure 3**: Trajectory of a 180-day rendezvous with asteroid 1991 VG including a 10-day stay

## IV. Planetary Flyby Missions

Venus, Earth, and Mars, the three terrestrial planets in the inner solar system with atmospheres, have a great deal in common for humans. Flyby missions to either Venus or Mars allow for proximate human study of Earth's planetary neighbors without the propellant and risk requirements associated with a rendezvous mission. We present detailed mission summaries in appendix C, and individual opportunities spanning the 2015-2025 decade in appendix D.

### A. Venus Flyby Mission

Planning for human missions to Venus began during the Apollo program[12], with a complete plan for a flyby expedition prepared by 1967. As for the Apollo lunar missions, this plan proposed to use the Saturn V rocket to launch an S-IVB Earth-departure stage and Command & Service Module (CSM) into orbit. After trans-Venus injection, the depleted S-IVB stage would be vented and converted for use as a crew habitat.

Flyby missions of Venus typically have a duration of one year and make use of a gravity assist to transfer the spacecraft into an Earth return trajectory. A baseline mission is presented in figure 4 for a December 2018 flyby. Such opportunities occur once every 1.6 years, the Earth-Venus synodic period. We find that Earth-escape $\Delta V$ from a 200 km LEO averages 3.72 km/s; this can be compared to the 3.15 km/s required for Apollo's trans-lunar injection[13]. The spacecraft will make its closest approach of Venus on the day side of the planet at an altitude on the order of 1,000 km, and will spend approximately three days within the planet's sphere of influence.

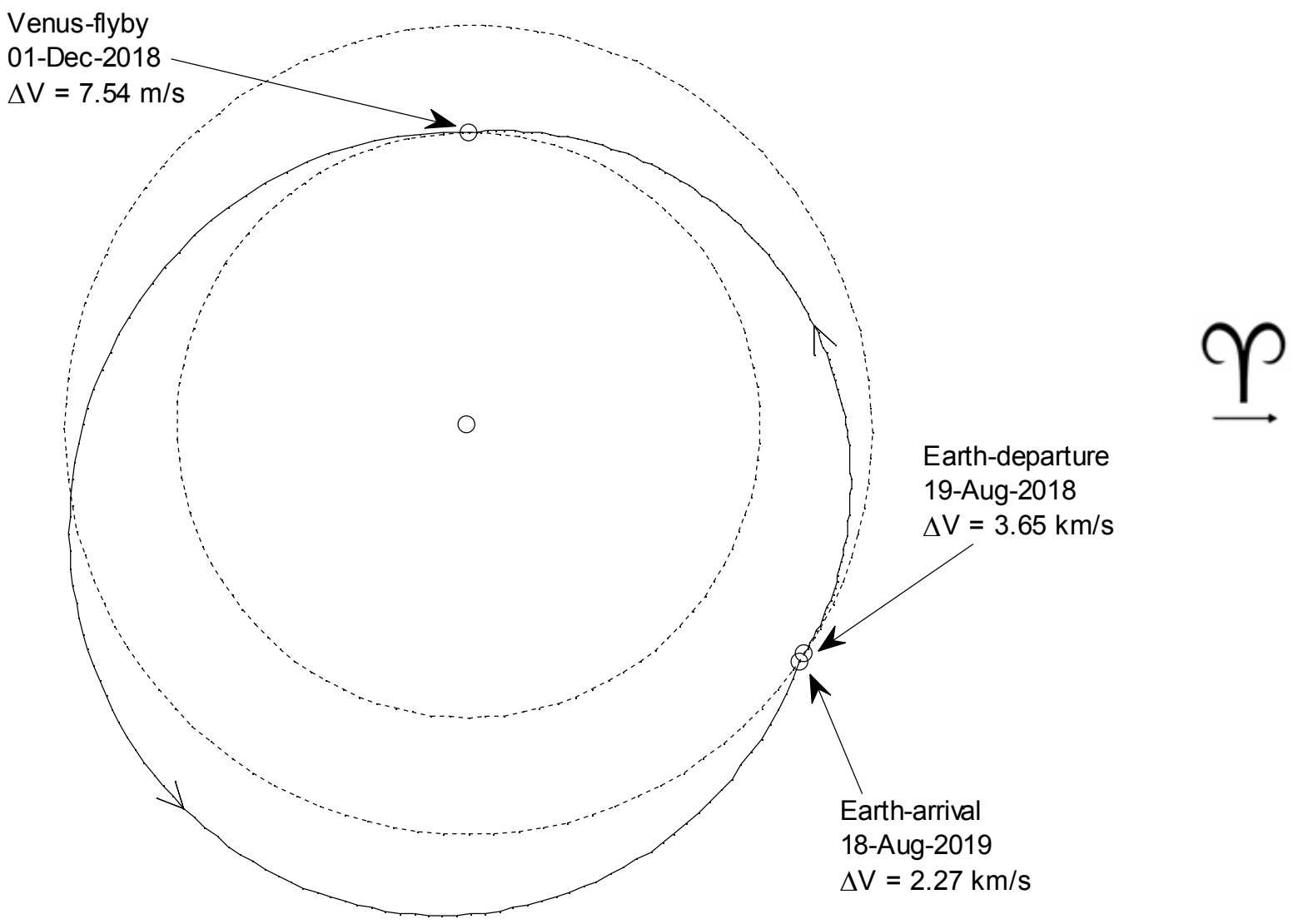

**Figure 4.** Trajectory of a one-year Venus flyby mission

### B. Mars Flyby Mission

While there are several ways to perform a Mars flyby[14], we find the lowest $\Delta V$ option is to launch into a 2-year period heliocentric orbit that intersects Mars on either the outbound or inbound leg. As a result, there are two opportunities seven months apart every 2.14-year Earth-Mars synodic period. Such missions will spend between one and two days within Mars' sphere of influence. The closest approach to Mars is on the order of 10,000 km and occurs near the dawn-dusk terminator. Figure 5 shows the trajectory for an outbound Mars flyby launching in May 2018. Injection into the 2-year flyby orbit requires on average 4.45 km/s $\Delta V$ from LEO.

The Mars flyby mission trajectory could also be crafted to flyby inner main-belt asteroids since the spacecraft will reach an aphelion of 2.17 AU.

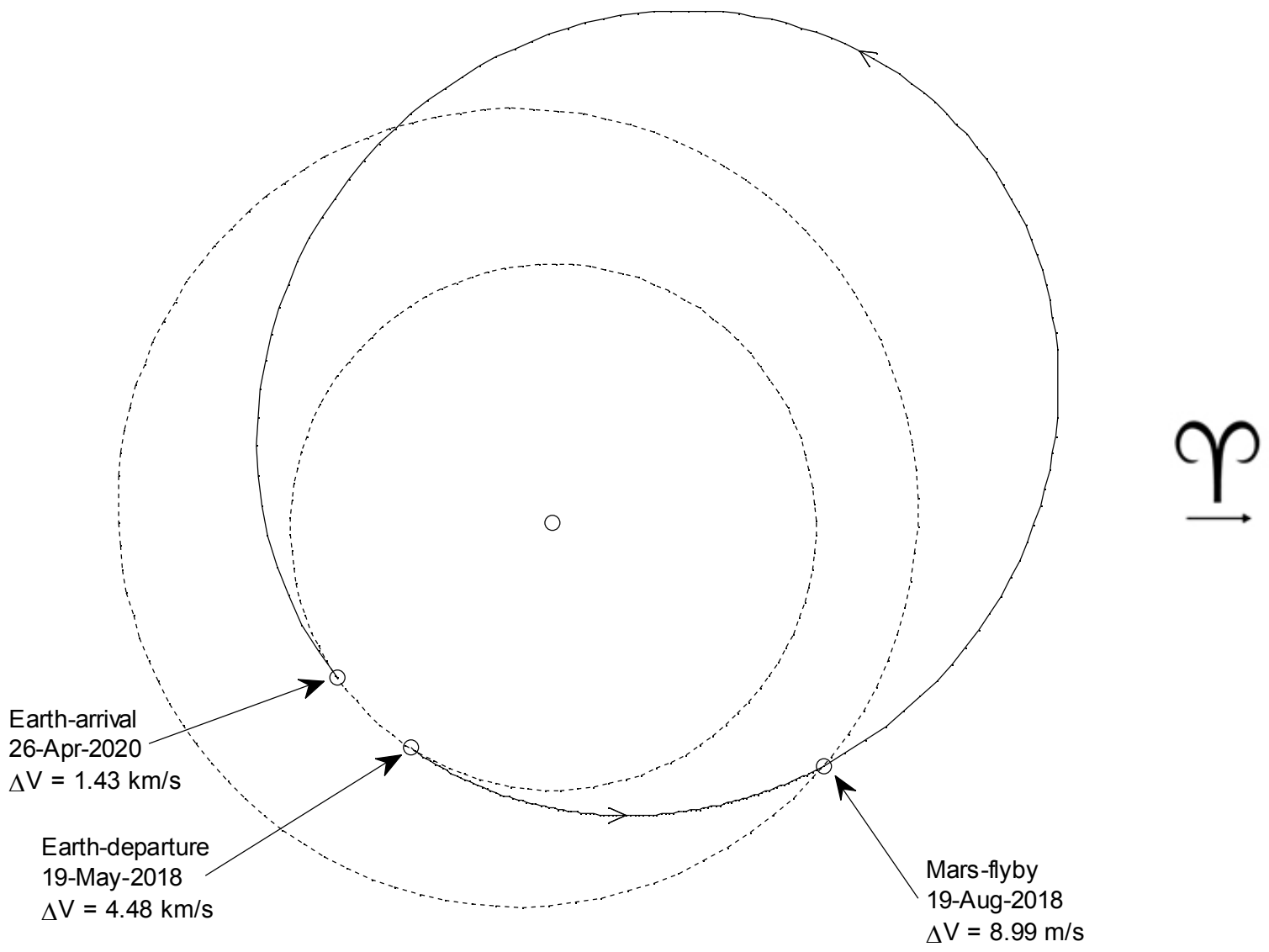

**Figure 5.** Trajectory of a two-year Mars flyby mission

### C. Venus-Mars Double Flyby Mission

We find a double flyby mission of Venus and Mars can take less time than a simple Mars flyby with similar ΔV requirements. Specifically, injection ΔV is on average 4.32 km/s while powered flyby maneuvers add up to 0.74 km/s. We find mission duration will be on average 1.44 years, with opportunities arising approximately every two years. Venus and Mars may be visited in either order depending on the opportunity, although missions which flyby Venus before Mars require less ΔV and tend to be shorter in duration. Figure 6 shows that the double flyby trajectory is similar to a Mars 1-month short-stay mission, but without the Mars stopover. Flyby altitudes are on the order of 500 km for each planet, typically lower than for simple flybys of either planet.

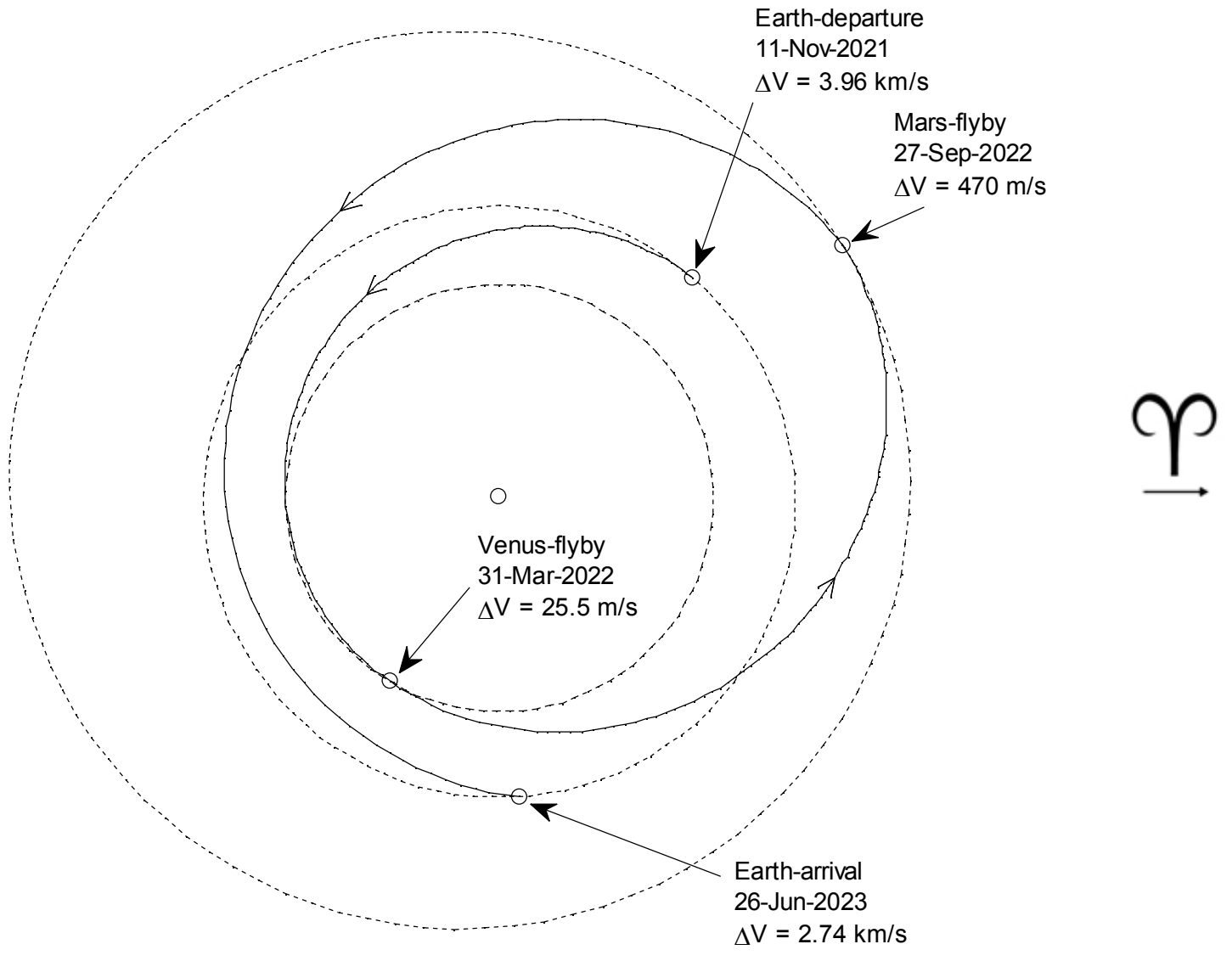

**Figure 6.** Trajectory of a Venus-Mars double flyby mission

## V. Planetary Rendezvous Missions

Planetary rendezvous missions present the most exciting human steps into the solar system in the foreseeable future. The most compelling targets besides near-Earth objects are the local gravity wells which may support eventual settlement and expansion by humans into the solar system.

### A. Planetary Staging Orbits

For rendezvous missions with Venus and Mars, it is of interest to enter a staging orbit about either planet with minimal aerocapture load on arrival and low ΔV requirement for Earth-return heliocentric injection. Likewise, during Earth-return it is beneficial to first enter a high-energy orbit through aerocapture to minimize thermal loads, followed by an aerobrake maneuver or direct atmospheric reentry depending on mission architecture.

Table 3 characterizes the high-energy staging orbits for Venus, Earth and Mars used in this paper. These orbits are defined to have a periapsis at the lowest altitude above any meaningful part of the planet's atmosphere[15], and an apoapsis such that the ΔV required to reach escape velocity is on the order of 100 m/s. A spacecraft can enter such a highly elliptical orbit after planetary aerocapture by making a small ΔV impulse at apoapsis, raising its periapsis outside of the planet's atmosphere. A spacecraft can then conduct its nominal mission from the staging orbit or adjust to a lower orbit (and in the case of Mars, visit Phobos or Deimos). For Earth-return heliocentric injection, a spacecraft takes maximum advantage of the Oberth effect by executing the injection burn at the staging orbit periapsis.

**Table 3.** Staging orbit parameters for Venus, Earth and Mars.

**Staging Orbit Parameters**

| | Periapsis (km) | Apoapsis (km) | Eccentricity | Altitude at periapsis (km) | Inertial speed at periapsis (km/s) | ΔV required for escape (m/s) | Orbital period (days) |
|---|---|---|---|---|---|---|---|
| **Venus** | 6,552 | 321,320 | 0.96 | 500 | 9.86 | 100 | 8.5 |
| **Earth** | 6,571 | 356,960 | 0.96 | 200 | 10.91 | 100 | 8.9 |
| **Mars** | 3,639 | 85,562 | 0.92 | 250 | 4.75 | 100 | 3.3 |

### B. Venus Rendezvous Mission

Two possible routes to Venus are presented in this paper: short-stay and long-stay rendezvous missions. The short-stay mission involves a 1-month stay in Venus orbit, a 1.18 year round-trip and 1.35 km/s post-Earth-escape ΔV. Figure 7 shows a typical trajectory of a short-stay mission, similar in nature to a Venus flyby trajectory but substituting the flyby for a 1-month stopover. Post-Earth-escape ΔV can be lowered by shortening the time spent in Venus orbit.

In contrast, the long-stay mission closely follows minimum energy Hohmann transfers. With aerobraking for arrival at Venus, this trajectory requires only 0.7 km/s post-Earth-departure ΔV, necessary for the Venus-departure burn. This mission also experiences lower aerocapture loads upon arrival at Venus. The long-stay trip requires 2 years total, with over one year in Venus orbit until planetary alignment for a return Hohmann transfer. As a result of mission geometry, Earth will pass through conjunction behind the sun as viewed by the crew in Venusian orbit.

Particularly exciting concepts for the scientific exploration of Venus in the near-term include a multiple-lander geophysical network; mobile surface exploration via airships or rovers; a sample return mission; and modern, high-resolution synthetic aperture radar (SAR) mapping[16]. We assume feasibility and technological limitations in the near-term will preclude a human landing on the surface of Venus. The most compelling missions to Venus in the foreseeable future, therefore, would not benefit substantially from a sustained human presence – as human explorers' roles would consist primarily of operating sophisticated remote-sensing equipment. Similarly, lander, airship, and balloon operations would benefit more from onboard automatic control than from human remote guidance and control.

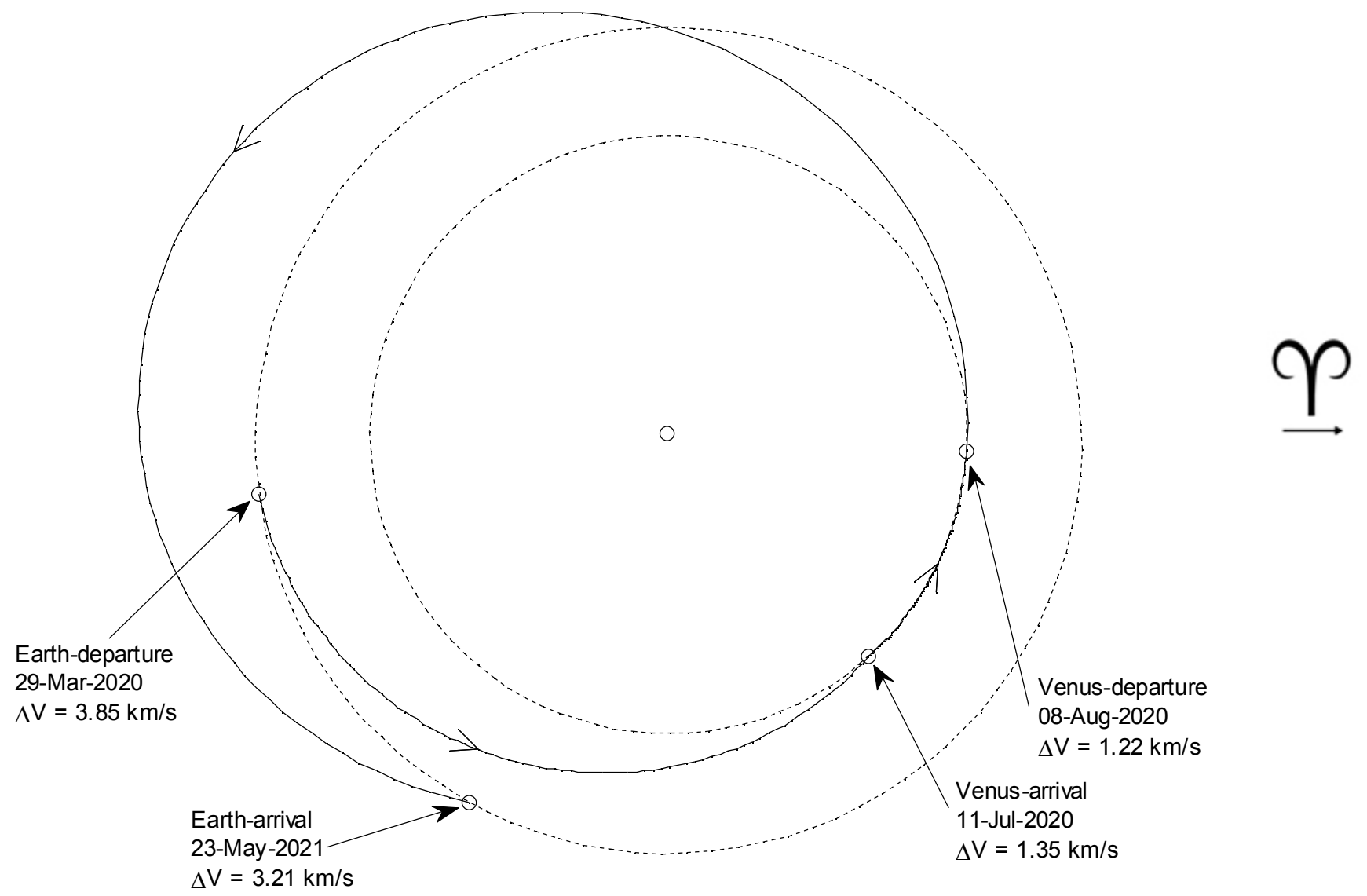

**Figure 7.** Trajectory of a short-stay Venus rendezvous mission spending 30 days in orbit.

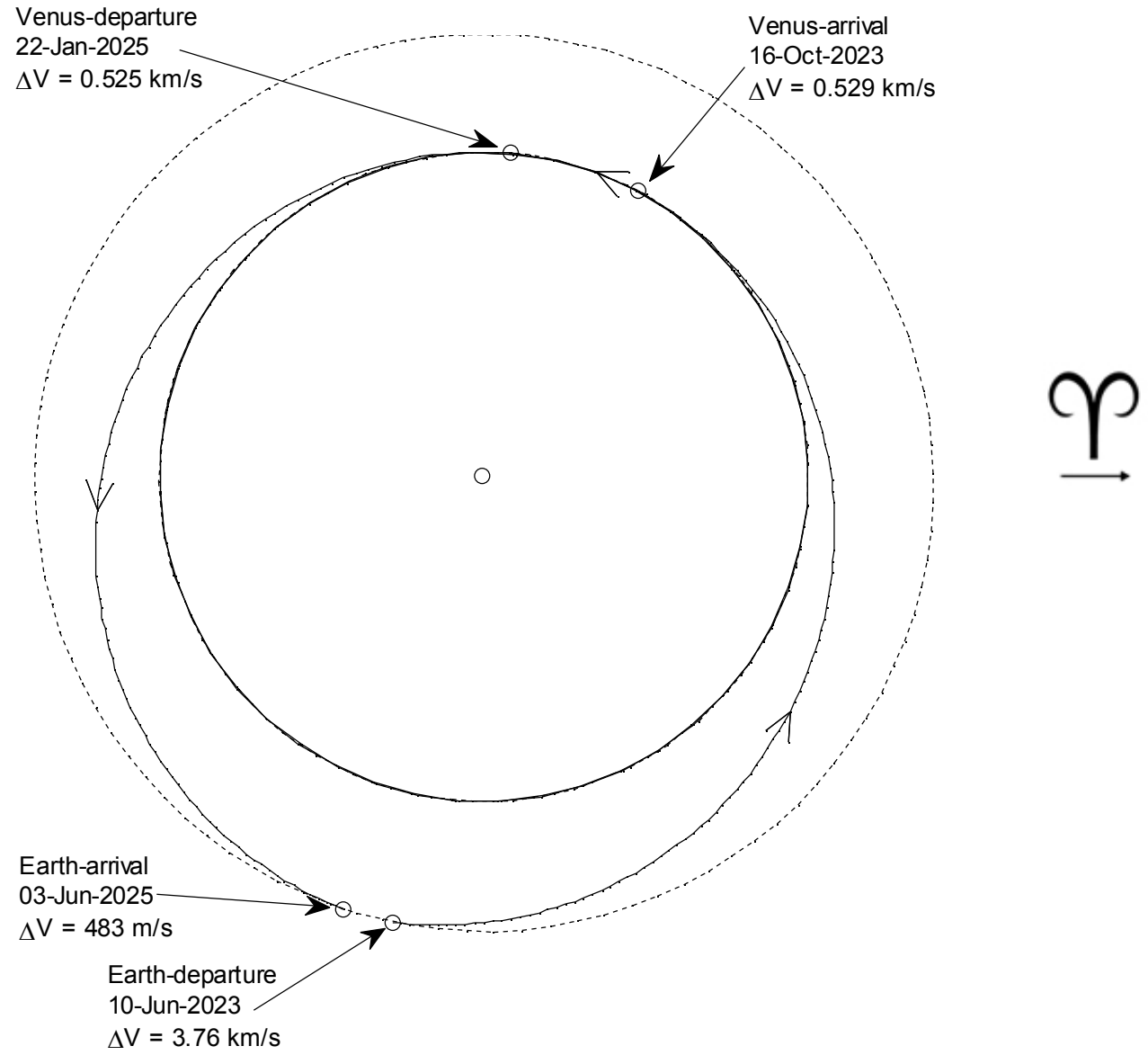

**Figure 8.** Trajectory of a long-stay Venus rendezvous mission. A spacecraft on such a mission would spend over one Venusian year in orbit before returning to Earth.

## C. Mars Rendezvous Mission

Short-stay and long-stay options are possible for Mars much like the Venus rendezvous missions. These missions are also known as opposition and conjunction missions, respectively, and have been discussed extensively in literature on human missions to explore the Martian surface[18]. Overviews are presented here for reference.

A Mars short-stay mission involves a 1-month orbital stay, which can be extended somewhat at the cost of higher $\Delta V$. Short-stay missions characteristically include a Venus flyby on either the outbound or inbound leg, depending on the relative position of Venus during a particular launch opportunity. Since the trajectory relies on a favorable alignment of three planets, post-Earth-escape ΔV and transit times vary considerably from year to year, but typically require a round-trip time of less than 1.6 years. A detailed breakdown of opportunities for the 2015-2025 decade may be viewed in appendix E.

The Mars long-stay rendezvous (conjunction-class) mission follows minimum-energy transfers allowing for over one Earth-year to be spent in Martian orbit. This trajectory also has lower ΔV and aerocapture requirements at the cost of a 2.5 year round-trip. Faster transit legs allow for more time to be spent in Mars orbit but require more ΔV.

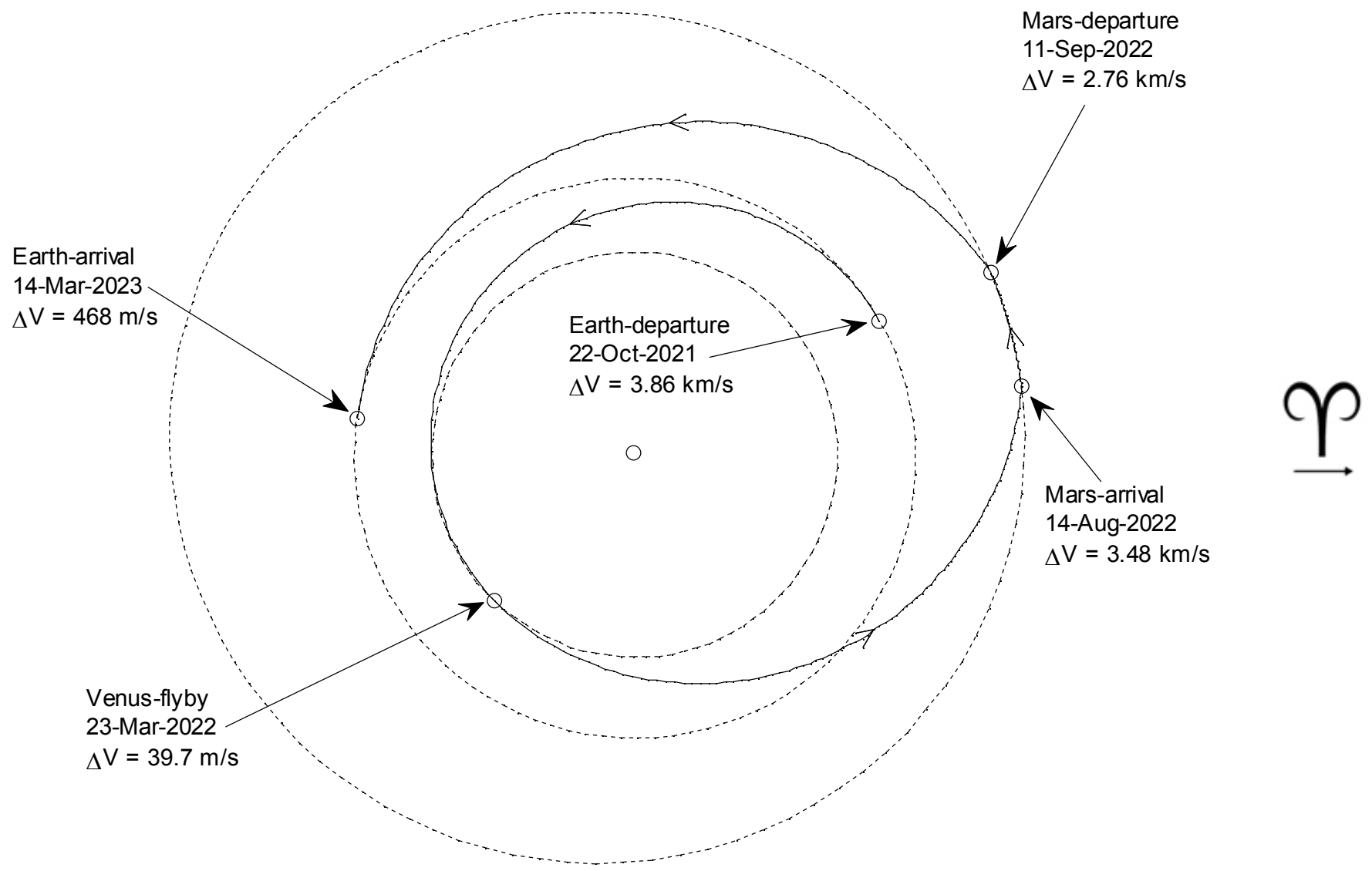

**Figure 9.** Trajectory of a short-stay Mars rendezvous mission

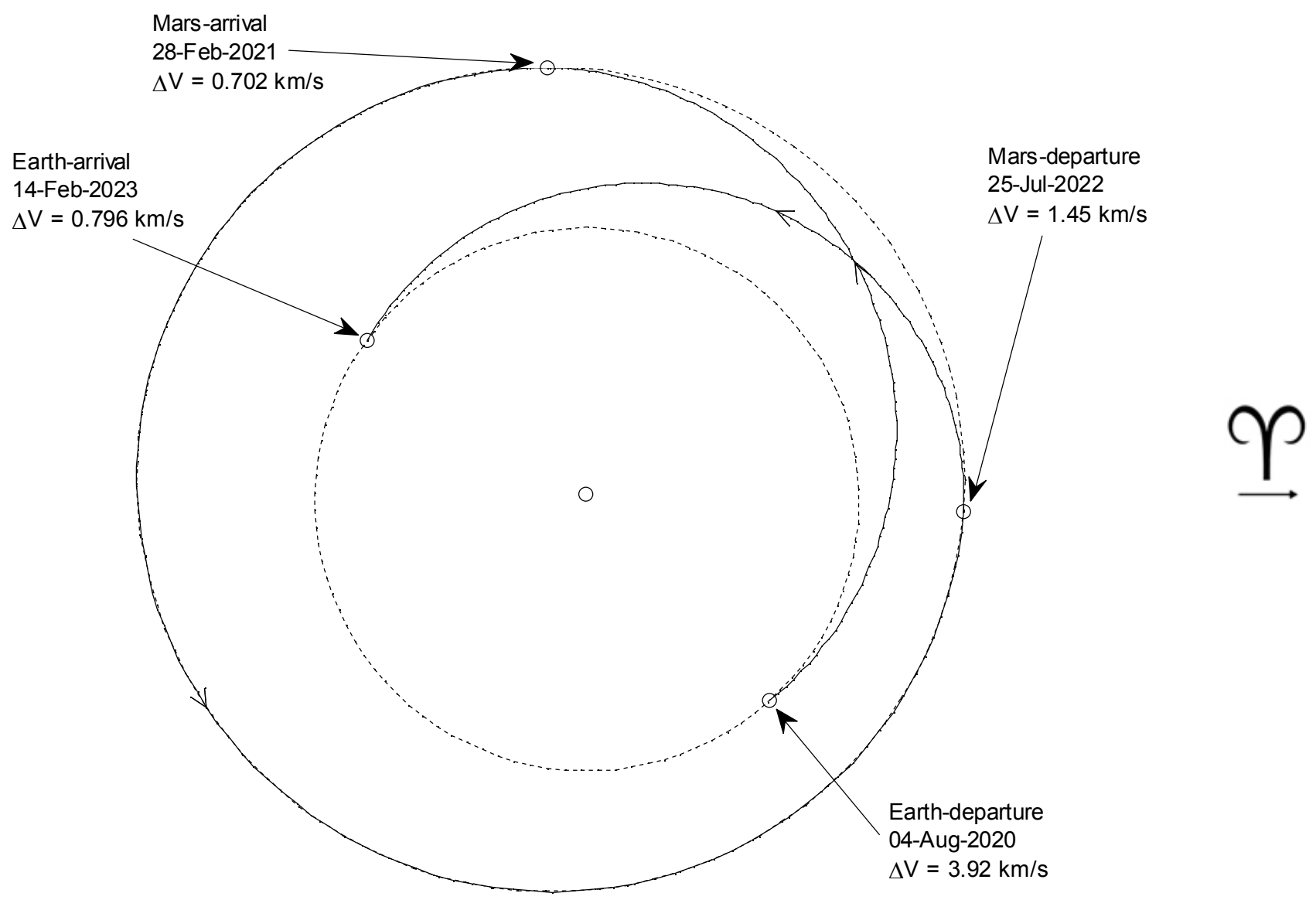

**Figure 10.** Trajectory of a long-stay Mars rendezvous mission

### D. Phobos and Deimos

Phobos and Deimos both possess near-equatorial and circular orbits around Mars with respective semi-major axes of 9,376 km and 23,458 km. Interest has been expressed in a rendezvous with Phobos as a precursor to landing humans on the surface of Mars; this would present lower ΔV requirements and remove risks associated with entry, descent, and landing (EDL) to the Martian surface on the first human expedition to the Martian system. Varyingly, discussion of Phobos as a rendezvous target has also focused on the possibility of teleoperated exploration of the Martian surface and a sample return from Phobos[19].

Table 4 indicates ΔV and transit time requirements to visit either one or both of the Martian moons from a high-energy staging orbit. The outer moon Deimos is the more accessible of the two and can be reached with a round-trip ΔV of 1.2 km/s using minimum-energy transfer arcs. A rendezvous with Phobos can take advantage of an aerobrake maneuver through the upper Martian atmosphere, resulting in a round-trip ΔV of 1.4 km/s. For visiting both Phobos and Deimos, it is best to rendezvous with Phobos first to take advantage of the available aerocapture maneuver; the spacecraft can then intercept Deimos through a Hohmann transfer, before returning to the staging orbit.

**Table 4.** Round-trip mission requirements from a Mars staging orbit to one or both of its moons

| Destination | Propulsive ΔV (km/s) | Aerobrake ΔV (km/s) | Transit Time (days) |
|---|---|---|---|
| **Deimos** | 1.2 | 0 | 4.4 |
| **Phobos** | 1.4 | 0.64 | 1.9 |
| **Phobos & Deimos** | 1.9 | 0.64 | 2.7 |

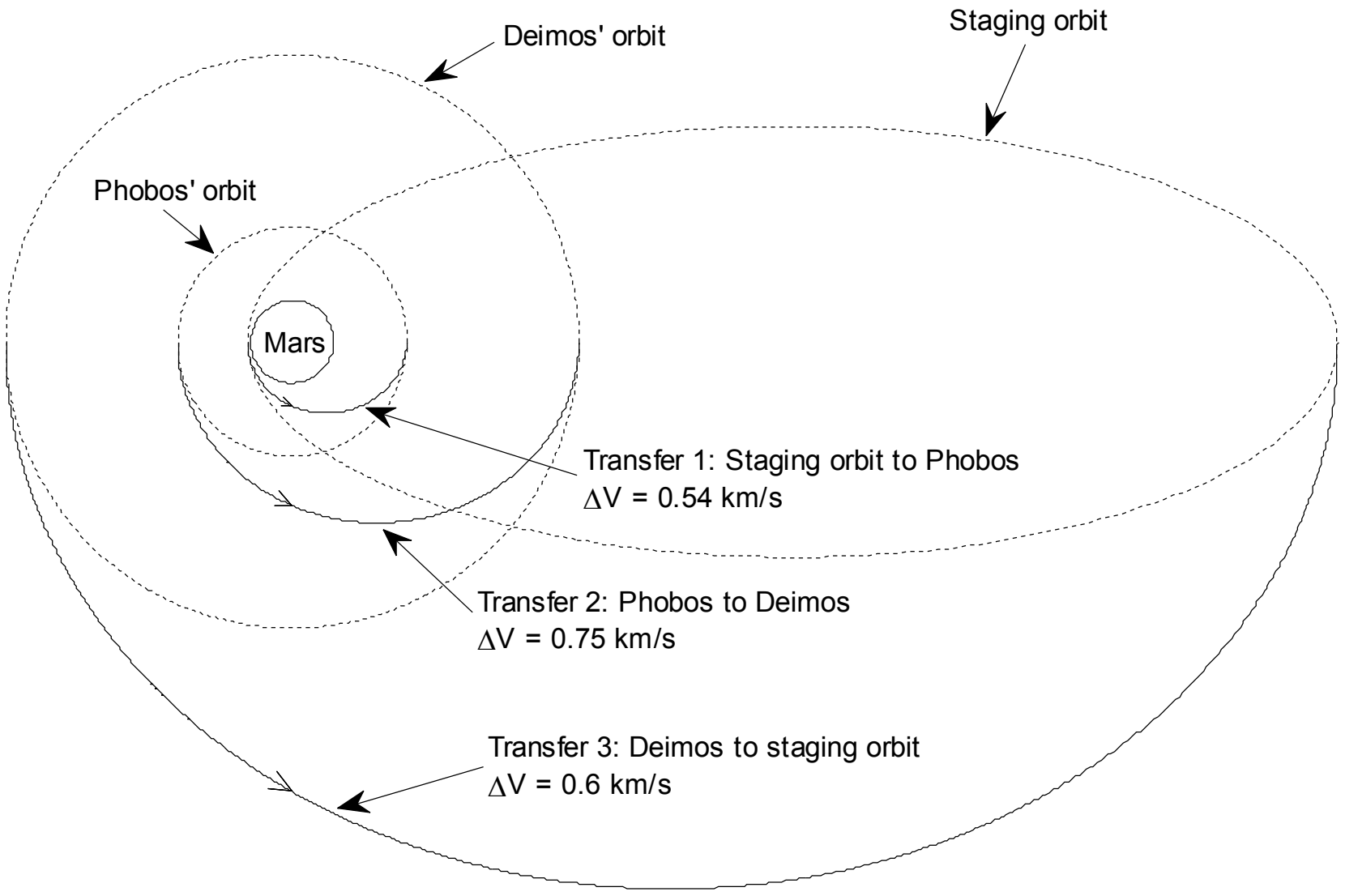

**Figure 11.** Round-trip trajectory from a Mars staging orbit to visit both Phobos and Deimos. An aerobrake maneuver reduces the staging orbit's apoapsis to intercept Phobos. After rendezvous with Phobos, the crew reaches Deimos via a Hohmann transfer, and subsequently returns to Mars staging orbit. In practice, the initial and final staging orbits will have different arguments of periapsides imposed by the arriving and departing hyperbolic asymptotes.

## VI. Conclusion

The flexible path, a term that came into widespread use during the Augustine Commission hearings, considered the possibility of sending humans beyond the Earth-Moon system as a new goal for human space exploration in the United States. Fundamental to any endeavor aiming to send humans beyond our planetary system will be solving the 'launch problem.' The cost and difficultly of access to space has been a uniquely-halting factor for human exploration of the solar system, and no single solution is currently at hand.

The diversification of actors who launch to and operate in space offers the prospect of sustainable, if gradual, reduction of the launch hurdle. The democratization of space is occurring for national governments, as more states pursue activities in orbit, and for individuals and non-state organizations, as small satellite capabilities expand. For NASA, we believe this will offer new partnerships, and necessitate new roles. The issues with subjecting long-term research and development plans to 2-year election cycles and annual budget redrafts are well known. If we see in human spaceflight a reflection of our ingenuity and accomplishments, we might see in political micromanagement of human spaceflight programmatics a particularly detrimental form of paternalism. This is not to motivate less Congressional attention to NASA, but a suggestion that other approaches to technology innovation, including those which build on the accomplishments of the commercial space sector, are likely to be necessary for achieving long-term goals in space.

NASA has a special role because, for 50 years, it has been a symbol for ingenuity and technological triumph – and a motivator for young people the way no other organization can. We should continue this tradition, and begin a program to send astronauts beyond the Earth-Moon system.* This is a timely issue for consideration as a more precise definition of the Augustine Commission's flexible path, and debate about the future direction of human spaceflight policy, take on programmatic urgency for NASA and the American space industry.

From the missions presented in this paper, we can begin to identify the requirements of an interplanetary spacecraft for missions within the inner solar system:

- 2.5-year reliability and life support systems appropriate for a human crew.
- Earth-departure C3 capability of less than 25 $km^2/s^2$; post-Earth-departure spacecraft $\Delta V$ capability of 4 km/s or less (within the capability of chemical propulsion); spacecraft aerocapture capability.
- Radiation and thermal systems capable of supporting humans to within 0.6 AU of the sun.
- Electrical power systems capable of supporting orbits to 2.2 AU from the sun.

*The authors hope to be among selectees for such a program.

## Appendix

### A. Examples of NEO Flyby and Rendezvous missions

| | Mission | Launch Date[a] | Mission Duration[b] (days) | Injection C3 ($km^2/s^2$) | Injection $\Delta V$[c] (km/s) | Post-escape $\Delta V$[d] (km/s) | Earth Aerocapture $\Delta V$[e] (km/s) | Earth Aerocapture Inertial (km/s) | Integrated Solar Exposure[f] ($kW\text{-}yr/m^2$) | Distance from Earth Max (AU) | Distance from Sun Min (AU) | Distance from Sun Max (AU) | Absolute Magnitude |
|---|---|---|---|---|---|---|---|---|---|---|---|---|---|
| | **Moon landing (Apollo)[13]** | Anytime | 12.5 | -1.7 | 3.15 | 6.40 | - | 11.0 | 0.05 | 0.003 | 1.00 | 1.00 | |
| Flyby | **171576 (1999 VP11)** | 4-Aug-2017 | 90 | 2.3 | 3.33 | 2.03 | 0.10 | 11.0 | 0.35 | 0.02 | 0.98 | 1.00 | 18.4 |
| Flyby | **163899 (2003 SD220)** | 8-Nov-2018 | 90 | 1.0 | 3.27 | 1.59 | 0.14 | 11.1 | 0.35 | 0.02 | 0.98 | 0.99 | 16.9 |
| Flyby | **7482 (1994 PC1)** | 5-Dec-2021 | 90 | 0.3 | 3.24 | 0.84 | 0.11 | 11.0 | 0.34 | 0.01 | 0.98 | 0.99 | 16.8 |
| Flyby | **45P/Honda-Mrkos-Pajdusakova[g]** | 16-Nov-2016 | 180 | 5.1 | 3.45 | 1.10 | 0.32 | 11.2 | 0.69 | 0.09 | 0.97 | 1.02 | 14.0 |
| Flyby | **144332 (2004 DV24)** | 4-Jul-2018 | 180 | 3.4 | 3.38 | 0.83 | 0.25 | 11.2 | 0.67 | 0.07 | 0.98 | 1.02 | 16.5 |
| Flyby | **66146 (1998 TU3)** | 9-Jul-2019 | 180 | 8.9 | 3.62 | 0.61 | 0.49 | 11.4 | 0.68 | 0.10 | 0.98 | 1.02 | 14.5 |
| Flyby | **21P/Giacobini-Zinner[g]** | 18-Mar-2018 | 365 | 9.5 | 3.65 | 0.04 | 0.53 | 11.4 | 1.37 | 0.39 | 0.92 | 1.08 | 10.7 |
| Flyby | **433 Eros** | 26-Nov-2018 | 365 | 15.3 | 3.90 | 1.08 | 1.04 | 12.0 | 1.37 | 0.58 | 0.86 | 1.15 | 11.2 |
| Flyby | **887 Alinda** | 6-Nov-2024 | 365 | 6.0 | 3.49 | 1.62 | 0.51 | 11.4 | 1.36 | 0.34 | 0.93 | 1.08 | 13.8 |
| Rendezvous | **2008 HU4** | 6-Apr-2016 | 90 | 1.3 | 3.29 | 2.56 | 0.60 | 11.5 | 0.33 | 0.02 | 1.00 | 1.02 | 28.2 |
| Rendezvous | **1991 VG** | 27-Dec-2017 | 90 | 5.5 | 3.47 | 5.13 | 1.88 | 12.8 | 0.35 | 0.05 | 0.98 | 1.00 | 28.4 |
| Rendezvous | **2008 EA9** | 2-Nov-2019 | 90 | 3.3 | 3.38 | 4.63 | 1.12 | 12.0 | 0.35 | 0.03 | 0.98 | 1.00 | 27.7 |
| Rendezvous | **2008 HU4** | 5-Apr-2016 | 180 | 1.2 | 3.28 | 1.98 | 0.44 | 11.3 | 0.65 | 0.03 | 1.00 | 1.03 | 28.2 |
| Rendezvous | **1991 VG** | 21-Jul-2017 | 180 | 2.9 | 3.36 | 2.29 | 1.77 | 12.7 | 0.67 | 0.08 | 0.97 | 1.03 | 28.4 |
| Rendezvous | **2008 EA9** | 28-Nov-2019 | 180 | 4.2 | 3.42 | 2.12 | 1.38 | 12.3 | 0.68 | 0.08 | 0.97 | 1.02 | 27.7 |
| Rendezvous | **2001 GP2** | 6-Nov-2019 | 365 | 2.9 | 3.36 | 1.57 | 1.12 | 12.0 | 1.37 | 0.20 | 0.96 | 1.04 | 26.9 |
| Rendezvous | **2007 UN12** | 22-May-2020 | 365 | 1.7 | 3.30 | 1.45 | 0.67 | 11.6 | 1.36 | 0.12 | 0.98 | 1.02 | 28.7 |
| Rendezvous | **99942 Apophis** | 6-May-2020 | 365 | 10.0 | 3.67 | 6.39 | 2.03 | 12.9 | 1.35 | 0.29 | 0.94 | 1.10 | 19.7 |

[a]Baseline solutions are presented. Departure date, duration, $\Delta V$ and aerocapture loads can be traded against one another to satisfy particular mission requirements.
[b]90-day, 180-day and 365-day rendezvous missions assume 5, 10 and 20-day stays respectively.
[c]$\Delta V$ required from a 200 km circular LEO to achieve injection C3
[d]Post-Earth-escape $\Delta V$ includes propulsive maneuvers required after injection, but does not include course corrections or margins.
[e]Aerocapture load for insertion into an Earth staging orbit. If aerocapture is not desired, a propulsive retro burn may be completed with approximately equivalent $\Delta V$.
[f]Integral of the solar radiation exposure throughout the trajectory, where the solar constant is assumed to be 1.372 $kW/m^2$ at 1 AU from the sun.
[g]Comet

## B. List of ΔV NEO rendezvous missions during the 2015-2025 decade

ΔV indicates propulsion requirements after injection from Earth orbit. All missions are constrained to have a C3 less than 15 $km^2/s^2$, ΔV less than 7 km/s and aerocapture inertial speed less than 13.4 km/s. We utilize fixed mission durations to provide a basis of comparison for mission requirements; in practice, a mission can be optimized to specific parameters of concern.

**90-day rendezvous missions with 5-day stay**

| Destination | Departure | ΔV-km/s |
|---|---|---|
| 2008 HU4 | 6-Apr-2016 | 2.56 |
| 2008 EA9 | 2-Nov-2019 | 4.63 |
| 2001 GP2 | 19-Sep-2020 | 5.03 |
| 1991 VG | 28-Dec-2017 | 5.13 |
| 2009 OS5 | 10-May-2020 | 5.56 |
| 2001 QJ142 | 10-Mar-2024 | 6.22 |
| 2007 UN12 | 4-Jun-2020 | 6.47 |
| 2007 YF | 21-Feb-2022 | 7.00 |

**180-day rendezvous missions with 10-day stay**

| Destination | Departure | ΔV-km/s |
|---|---|---|
| 2008 HU4 | 6-Apr-2016 | 1.97 |
| 2008 EA9 | 28-Nov-2019 | 2.12 |
| 1991 VG | 21-Jul-2017 | 2.29 |
| 2009 OS5 | 5-Mar-2020 | 3.70 |
| 2007 UN12 | 20-Jun-2020 | 3.95 |
| 2003 WT153 | 30-Sep-2019 | 4.46 |
| 2003 YS70 | 6-Dec-2022 | 4.66 |
| 2001 GP2 | 6-May-2020 | 4.77 |
| 2001 QJ142 | 31-Jan-2024 | 4.78 |
| 2004 QA22 | 31-May-2017 | 4.85 |
| 2008 UD95 | 11-May-2019 | 4.94 |
| 2001 BB16 | 23-Dec-2015 | 4.99 |
| 2005 KA | 7-Apr-2015 | 5.02 |
| 2010 FY9 | 8-Oct-2020 | 5.24 |
| 2009 QR | 14-Aug-2023 | 5.32 |
| 2004 VJ1 | 19-Oct-2015 | 5.50 |
| 2007 RF1 | 20-Jul-2020 | 5.68 |
| 2006 WB | 17-Jun-2024 | 5.71 |
| 2005 UV64 | 29-Sep-2020 | 5.76 |
| 2003 CA4 | 15-Jan-2018 | 5.93 |
| 2007 YF | 26-Jan-2022 | 5.94 |
| 2007 BB | 8-Jul-2024 | 6.16 |
| 2006 QV89 | 13-Sep-2019 | 6.39 |
| 2004 JN1 | 28-Sep-2020 | 6.43 |
| 162173 | 16-Nov-2020 | 6.57 |
| 2005 OH3 | 7-Mar-2016 | 6.69 |
| 2008 XS | 3-Oct-2021 | 6.74 |
| 2000 SZ162 | 17-Apr-2017 | 6.97 |

**365-day rendezvous missions with 20-day stay**

| Destination | Departure | ΔV-km/s |
|---|---|---|
| 1991 VG | 5-Jun-2017 | 1.01 |
| 2001 GP2 | 28-Sep-2020 | 1.32 |
| 2007 UN12 | 22-May-2020 | 1.45 |
| 2008 EA9 | 24-Oct-2019 | 1.55 |
| 2008 HU4 | 29-Mar-2016 | 1.99 |
| 2009 YR | 27-Jul-2019 | 2.27 |
| 2009 BD | 29-Oct-2022 | 2.38 |
| 2009 DB43 | 1-Sep-2015 | 2.87 |
| 2008 DL4 | 29-Aug-2016 | 2.99 |
| 2009 YF | 7-May-2019 | 3.12 |
| 2004 FM32 | 7-Oct-2018 | 3.14 |
| 2007 YF | 1-Jun-2021 | 3.23 |
| 2008 GM2 | 17-Oct-2020 | 3.23 |
| 2001 CQ36 | 25-Jul-2021 | 3.24 |
| 2005 TG50 | 19-Feb-2021 | 3.31 |
| 2007 UY1 | 25-Mar-2021 | 3.46 |
| 2009 CV | 6-Sep-2015 | 3.54 |
| 2007 BB | 18-Feb-2015 | 3.62 |
| 2008 UD95 | 12-Nov-2018 | 3.74 |
| 2001 QJ142 | 20-Apr-2024 | 3.83 |
| 2006 FH36 | 18-Sep-2020 | 3.83 |
| 2004 QA22 | 17-Sep-2017 | 3.85 |
| 1999 CG9 | 29-Jul-2021 | 3.87 |
| 2009 OS5 | 27-Feb-2020 | 3.99 |
| 2007 HL4 | 4-Nov-2019 | 4.10 |
| 1999 VX25 | 10-Jul-2022 | 4.14 |
| 2006 UB17 | 22-Apr-2017 | 4.17 |
| 1999 AO10 | 9-Nov-2018 | 4.37 |
| 2000 SZ162 | 10-Jul-2017 | 4.37 |
| 2000 LG6 | 7-Dec-2021 | 4.48 |
| 2007 RO17 | 4-Feb-2024 | 4.54 |
| 2008 CX118 | 18-Sep-2023 | 4.56 |
| 2001 BA16 | 3-Jul-2021 | 4.63 |
| 2004 JN1 | 11-Sep-2020 | 4.65 |
| 2006 HE2 | 26-Nov-2017 | 4.68 |
| 2010 LG61 | 23-Apr-2018 | 4.74 |
| 2009 RT1 | 26-Feb-2019 | 4.76 |
| 2004 WH1 | 17-May-2021 | 4.80 |
| 2003 CA4 | 16-Aug-2017 | 4.81 |
| 2005 UE1 | 9-May-2021 | 4.81 |
| 2003 WT153 | 21-May-2019 | 4.87 |
| 2010 AN61 | 15-Jul-2019 | 4.88 |
| 2001 BB16 | 18-Dec-2015 | 4.93 |
| 2010 FY9 | 22-Apr-2020 | 4.98 |
| 2006 UQ216 | 16-Jun-2021 | 5.00 |
| 2005 UV64 | 29-Apr-2020 | 5.04 |
| 2006 XP4 | 28-Sep-2019 | 5.08 |
| 2004 VJ1 | 11-Feb-2015 | 5.09 |
| 2001 QC34 | 6-Jun-2019 | 5.18 |
| 2004 EU22 | 30-Sep-2017 | 5.22 |
| 2003 LN6 | 17-Mar-2022 | 5.24 |
| 162173 | 21-May-2020 | 5.27 |
| 2005 KA | 8-Apr-2015 | 5.28 |
| 2007 UD6 | 13-May-2018 | 5.30 |
| 2009 QR | 14-Aug-2023 | 5.32 |
| 2006 QV89 | 16-Jul-2019 | 5.33 |
| 2007 TF15 | 11-Dec-2020 | 5.38 |
| 2005 VL1 | 22-Apr-2016 | 5.38 |
| 1998 KG3 | 4-Nov-2017 | 5.40 |
| 2008 BT2 | 9-May-2021 | 5.42 |
| 2010 KV7 | 11-Dec-2017 | 5.45 |
| 2010 DJ | 24-Jul-2017 | 5.52 |
| 2010 GV23 | 2-Oct-2018 | 5.56 |
| 2001 KM20 | 19-Nov-2018 | 5.57 |
| 2007 RF1 | 20-Jul-2020 | 5.59 |
| 2005 FG | 13-Oct-2017 | 5.63 |
| 2004 BY1 | 28-Jun-2023 | 5.66 |
| 2000 SJ344 | 10-Apr-2017 | 5.73 |
| 2008 WY94 | 4-Jun-2022 | 5.76 |
| 2005 VN5 | 3-Jan-2016 | 5.84 |
| 2005 OH3 | 25-Aug-2015 | 5.85 |
| 2004 OW10 | 14-Feb-2023 | 5.95 |
| 2009 YS | 25-Jun-2022 | 5.95 |
| 2003 YS70 | 31-Oct-2022 | 5.96 |
| 2005 QP87 | 13-Apr-2016 | 6.02 |
| 2006 BU7 | 29-Jul-2019 | 6.02 |
| 2002 TS69 | 15-Apr-2016 | 6.04 |
| 2006 SY5 | 7-Sep-2022 | 6.12 |
| 2008 LG2 | 14-Feb-2019 | 6.21 |
| 2004 UT1 | 6-Dec-2021 | 6.24 |
| 2007 RC20 | 22-Oct-2020 | 6.29 |
| 2008 EV5 | 21-Nov-2023 | 6.32 |
| 2006 BA | 29-Jul-2018 | 6.36 |
| 2005 TA | 2-Apr-2018 | 6.37 |
| 99942 Apophis | 26-May-2020 | 6.39 |
| 2005 TH50 | 26-Jul-2018 | 6.40 |
| 2006 XW | 13-Nov-2020 | 6.40 |
| 2004 FN8 | 28-Sep-2022 | 6.41 |
| 2008 JP24 | 27-Nov-2021 | 6.41 |
| 2009 BF2 | 9-Aug-2019 | 6.51 |
| 2005 GN22 | 28-Sep-2020 | 6.64 |
| 2007 PS9 | 6-Feb-2017 | 6.66 |
| 2009 TD8 | 21-Dec-2018 | 6.67 |
| 2005 CD69 | 12-Sep-2021 | 6.68 |
| 2004 XK3 | 4-May-2023 | 6.70 |
| 2007 HB15 | 29-Oct-2020 | 6.78 |
| 2004 SR | 28-Mar-2020 | 6.79 |
| 2009 TD17 | 7-Jan-2021 | 6.82 |
| 2008 TS10 | 24-Mar-2015 | 6.82 |
| 1999 SH10 | 13-Oct-2021 | 6.83 |
| 2008 VC | 1-Jun-2021 | 6.84 |
| 2006 HC | 22-Nov-2019 | 6.84 |
| 1999 VW25 | 23-May-2016 | 6.85 |
| 2009 WC106 | 12-May-2018 | 6.86 |
| 2008 XS | 9-Jun-2021 | 6.90 |
| 2009 UZ87 | 20-Aug-2017 | 6.90 |
| 2006 WB | 2-Jun-2024 | 6.98 |

### C. Overview of planetary missions

Values indicate averages over missions spanning the 2015-2025 decade.

| Mission | Opportunity Frequency (years) | Mission Duration (years) | Time spent at destination | Injection | | Post-escape ΔV[a] (km/s) | Destination Aerocapture[b] | | Earth Aerocapture | | Integrated Solar Exposure (kW-yr/$m^2$) | Distance from Earth | Distance from Sun | |
|---|---|---|---|---|---|---|---|---|---|---|---|---|---|---|
| | | | | C3 ($km^2/s^2$) | ΔV (km/s) | | ΔV (km/s) | Inertial (km/s) | ΔV (km/s) | Inertial (km/s) | | Max (AU) | Min (AU) | Max (AU) |
| **Venus flyby** | 1.60 | 1.01 | N/A | 11.2 | 3.72 | 0.02 | N/A | N/A | 2.54 | 13.5 | 1.49 | 0.82 | 0.72 | 1.21 |
| **Mars flyby** | 1.08 | 1.95 | N/A | 28.4 | 4.45 | 0.02 | N/A | N/A | 1.33 | 12.2 | 1.12 | 3.17 | 1.00 | 2.17 |
| **Venus-Mars flybys** | 1.85 | 1.44 | N/A | 25.3 | 4.32 | 0.74 | N/A | N/A | 2.27 | 13.2 | 2.13 | 1.64 | 0.59 | 1.48 |
| **Venus short-stay** | 1.60 | 1.18 | 1 month | 10.8 | 3.71 | 1.35 | 1.08 | 10.9 | 3.10 | 14.0 | 1.74 | 1.15 | 0.72 | 1.30 |
| **Venus long-stay** | 1.60 | 1.94 | 1.3 years | 11.0 | 3.71 | 0.71 | 0.88 | 10.7 | 0.76 | 11.7 | 4.65 | 1.72 | 0.72 | 1.00 |
| **Mars short-stay** | 1.66 | 1.49 | 1 month | 16.2 | 3.94 | 2.30 | 2.72 | 7.5 | 1.19 | 12.1 | 2.13 | 1.53 | 0.63 | 1.49 |
| **Mars long-stay** | 2.14 | 2.57 | 1.2 years | 13.0 | 3.80 | 1.02 | 0.92 | 5.7 | 0.62 | 11.5 | 1.89 | 2.59 | 0.99 | 1.66 |

See relevant footnotes from appendix A

[a]For rendezvous missions, ΔV does not include maneuvers conducted beyond insertion into a staging orbit (e.g. a lower Venus orbit or a rendezvous with Phobos and Deimos).

### D. List of planetary flyby opportunities during the 2015-2025 decade.

| | Departure Date | Mission Duration (years) | Injection | | Post-escape ΔV (km/s) | Earth Aerocapture | | Integrated Solar Exposure (kW-yr/$m^2$) | Distance from Earth | Distance from Sun | | Type |
|---|---|---|---|---|---|---|---|---|---|---|---|---|
| | | | C3 ($km^2/s^2$) | ΔV (km/s) | | ΔV (km/s) | Inertial (km/s) | | Max (AU) | Min (AU) | Max (AU) | |
| **Venus Flyby** | 27-May-2015 | 1.01 | 11.5 | 3.74 | 0.03 | 2.54 | 13.5 | 1.48 | 0.80 | 0.73 | 1.20 | |
| | 31-Dec-2016 | 1.03 | 9.6 | 3.65 | 0.06 | 2.58 | 13.5 | 1.52 | 0.82 | 0.72 | 1.21 | |
| | 19-Aug-2018 | 1.00 | 9.6 | 3.65 | 0.01 | 2.27 | 13.2 | 1.46 | 0.80 | 0.72 | 1.22 | |
| | 25-Mar-2020 | 1.01 | 13.1 | 3.80 | 0.00 | 2.78 | 13.7 | 1.48 | 0.87 | 0.73 | 1.21 | |
| | 26-Oct-2021 | 1.01 | 13.3 | 3.81 | 0.01 | 2.51 | 13.4 | 1.48 | 0.79 | 0.72 | 1.22 | |
| | 25-May-2023 | 1.01 | 11.7 | 3.75 | 0.01 | 2.56 | 13.5 | 1.48 | 0.81 | 0.73 | 1.20 | |
| | 29-Dec-2024 | 1.03 | 9.7 | 3.66 | 0.03 | 2.58 | 13.5 | 1.52 | 0.82 | 0.72 | 1.21 | |
| **Mars Flyby** | 10-Mar-2016 | 1.94 | 27.6 | 4.41 | 0.01 | 1.37 | 12.3 | 1.10 | 3.17 | 0.99 | 2.18 | Outbound Flyby |
| | 20-Oct-2016 | 1.96 | 27.6 | 4.41 | 0.02 | 1.28 | 12.2 | 1.13 | 3.17 | 1.00 | 2.17 | Inbound Flyby |
| | 19-May-2018 | 1.94 | 29.1 | 4.48 | 0.01 | 1.43 | 12.3 | 1.10 | 3.17 | 1.01 | 2.16 | Outbound Flyby |
| | 10-Jan-2019 | 1.95 | 27.9 | 4.43 | 0.03 | 1.27 | 12.2 | 1.12 | 3.17 | 0.98 | 2.19 | Inbound Flyby |
| | 4-Aug-2020 | 1.94 | 28.4 | 4.45 | 0.01 | 1.38 | 12.3 | 1.10 | 3.17 | 1.01 | 2.16 | Outbound Flyby |
| | 8-Mar-2021 | 1.95 | 26.5 | 4.37 | 0.02 | 1.30 | 12.2 | 1.12 | 3.17 | 0.99 | 2.18 | Inbound Flyby |
| | 15-Sep-2022 | 1.96 | 26.9 | 4.39 | 0.10 | 1.38 | 12.3 | 1.13 | 3.17 | 1.01 | 2.17 | Outbound Flyby |
| | 27-Apr-2023 | 1.97 | 28.9 | 4.47 | 0.01 | 1.25 | 12.2 | 1.14 | 3.17 | 1.00 | 2.17 | Inbound Flyby |
| | 3-Nov-2024 | 1.95 | 32.5 | 4.61 | 0.01 | 1.27 | 12.2 | 1.12 | 3.17 | 0.99 | 2.17 | Outbound Flyby |
| **Venus-Mars Flybys** | 11-Apr-2016 | 1.17 | 35.7 | 4.74 | 1.56 | 1.55 | 12.5 | 1.75 | 1.35 | 0.59 | 1.40 | Mars then Venus |
| | 6-Apr-2017 | 1.35 | 22.3 | 4.19 | 0.10 | 2.88 | 13.8 | 2.04 | 1.89 | 0.56 | 1.54 | Venus then Mars |
| | 28-May-2020 | 1.69 | 26.7 | 4.38 | 1.45 | 1.04 | 12.0 | 2.46 | 1.82 | 0.57 | 1.53 | Mars then Venus |
| | 11-Nov-2021 | 1.62 | 16.7 | 3.96 | 0.50 | 2.74 | 13.7 | 2.27 | 1.27 | 0.71 | 1.44 | Venus then Mars |
| | 10-Sep-2023 | 1.37 | 25.0 | 4.31 | 0.07 | 3.16 | 14.1 | 2.14 | 1.89 | 0.53 | 1.49 | Venus then Mars |

See relevant footnotes from appendix A

## E. List of planetary rendezvous opportunities during the 2015-2025 decade.

| | Departure Date | Mission Duration (years) | Time spent at destination | Injection | | Post-escape ΔV (km/s) | Destination Aerocapture | | Earth Aerocapture | | Integrated Solar Exposure (kW-yr/m$^2$) | Distance from Earth | Distance from Sun | | Venus Flyby Leg |
|---|---|---|---|---|---|---|---|---|---|---|---|---|---|---|---|
| | | | | C3 (km$^2$/s$^2$) | ΔV (km/s) | | ΔV (km/s) | Inertial (km/s) | ΔV (km/s) | Inertial (km/s) | | Max (AU) | Min (AU) | Max (AU) | |
| Venus short-stay | 4-Jun-2015 | 1.17 | 1 month | 11.8 | 3.75 | 1.42 | 0.95 | 10.8 | 2.87 | 13.8 | 1.72 | 1.12 | 0.72 | 1.30 | |
| | 31-Dec-2016 | 1.23 | 1 month | 7.8 | 3.58 | 1.42 | 0.89 | 10.8 | 3.37 | 14.3 | 1.81 | 1.22 | 0.73 | 1.30 | |
| | 11-Aug-2018 | 1.18 | 1 month | 8.0 | 3.59 | 1.33 | 1.31 | 11.2 | 2.92 | 13.8 | 1.74 | 1.09 | 0.72 | 1.30 | |
| | 29-Mar-2020 | 1.15 | 1 month | 14.1 | 3.85 | 1.22 | 1.35 | 11.2 | 3.21 | 14.1 | 1.69 | 1.16 | 0.73 | 1.29 | |
| | 3-Nov-2021 | 1.14 | 1 month | 14.0 | 3.85 | 1.22 | 1.22 | 11.1 | 3.11 | 14.0 | 1.68 | 1.09 | 0.72 | 1.29 | |
| | 2-Jun-2023 | 1.17 | 1 month | 12.1 | 3.76 | 1.41 | 0.94 | 10.8 | 2.88 | 13.8 | 1.72 | 1.12 | 0.72 | 1.30 | |
| | 29-Dec-2024 | 1.23 | 1 month | 7.9 | 3.58 | 1.42 | 0.86 | 10.7 | 3.37 | 14.3 | 1.81 | 1.22 | 0.73 | 1.30 | |
| Venus long-stay | 12-Jun-2015 | 1.97 | 1.27 years | 11.5 | 3.74 | 0.54 | 0.54 | 10.4 | 0.52 | 11.4 | 4.68 | 1.74 | 0.72 | 1.02 | |
| | 12-Jan-2017 | 1.94 | 1.30 years | 8.4 | 3.60 | 0.53 | 0.85 | 10.7 | 0.91 | 11.8 | 4.64 | 1.71 | 0.72 | 0.98 | |
| | 23-Aug-2018 | 1.93 | 1.35 years | 10.9 | 3.71 | 0.86 | 1.10 | 11.0 | 1.11 | 12.0 | 4.63 | 1.73 | 0.72 | 1.02 | |
| | 29-Mar-2020 | 1.93 | 1.35 years | 15.8 | 3.92 | 1.01 | 1.11 | 11.0 | 0.85 | 11.8 | 4.63 | 1.72 | 0.72 | 1.00 | |
| | 15-Nov-2021 | 1.93 | 1.20 years | 9.6 | 3.65 | 0.95 | 1.20 | 11.1 | 0.57 | 11.5 | 4.62 | 1.72 | 0.70 | 1.00 | |
| | 10-Jun-2023 | 1.98 | 1.27 years | 12.1 | 3.76 | 0.53 | 0.53 | 10.4 | 0.48 | 11.4 | 4.70 | 1.74 | 0.72 | 1.02 | |
| | 10-Jan-2025 | 1.94 | 1.30 years | 8.4 | 3.60 | 0.52 | 0.82 | 10.7 | 0.89 | 11.8 | 4.64 | 1.71 | 0.72 | 0.98 | |
| Mars short-stay | 27-May-2015 | 1.63 | 1 month | 13.2 | 3.81 | 2.29 | 3.22 | 8.0 | 3.42 | 14.3 | 2.26 | 1.12 | 0.72 | 1.55 | Outbound |
| | 27-Feb-2016 | 1.25 | 1 month | 9.2 | 3.64 | 3.89 | 2.34 | 7.1 | 1.31 | 12.2 | 1.78 | 1.27 | 0.62 | 1.41 | Inbound |
| | 29-Mar-2017 | 1.54 | 1 month | 20.1 | 4.10 | 1.04 | 2.61 | 7.4 | 0.58 | 11.5 | 2.23 | 1.86 | 0.58 | 1.55 | Outbound |
| | 19-Jul-2020 | 1.50 | 1 month | 14.5 | 3.87 | 2.80 | 1.41 | 6.2 | 0.86 | 11.8 | 2.15 | 1.78 | 0.61 | 1.54 | Inbound |
| | 22-Oct-2021 | 1.39 | 1 month | 14.4 | 3.86 | 2.80 | 3.48 | 8.2 | 0.47 | 11.4 | 1.94 | 1.24 | 0.72 | 1.43 | Outbound |
| | 14-Sep-2023 | 1.64 | 1 month | 25.7 | 4.34 | 0.99 | 3.29 | 8.0 | 0.48 | 11.4 | 2.43 | 1.89 | 0.53 | 1.44 | Outbound |
| Mars long-stay | 18-Mar-2016 | 2.54 | 1.41 years | 14.1 | 3.85 | 0.70 | 1.34 | 6.1 | 0.62 | 11.5 | 1.83 | 2.66 | 1.00 | 1.67 | |
| | 23-May-2018 | 2.55 | 1.48 years | 8.6 | 3.61 | 1.17 | 0.95 | 5.7 | 0.62 | 11.5 | 1.85 | 2.68 | 0.98 | 1.67 | |
| | 4-Aug-2020 | 2.53 | 1.40 years | 15.7 | 3.92 | 1.45 | 0.70 | 5.5 | 0.80 | 11.7 | 1.83 | 2.64 | 0.99 | 1.67 | |
| | 15-Sep-2022 | 2.63 | 0.85 years[a] | 14.4 | 3.86 | 0.94 | 0.88 | 5.6 | 0.50 | 11.4 | 1.97 | 2.55 | 1.01 | 1.69 | |
| | 14-Oct-2024 | 2.62 | 0.87 years[a] | 12.1 | 3.76 | 0.82 | 0.73 | 5.5 | 0.58 | 11.5 | 1.95 | 2.42 | 1.00 | 1.62 | |

See relevant footnotes from appendix A

[a]Low ΔV route follows a type II trajectory resulting in a shorter stay at Mars

## Acknowledgments

The authors would like to thank the NASA Ames Research Center and the Ames Mission Design Center for providing the tools, insights, and opportunity to work on this paper. NASA has provided the opportunities for our education and work in the field of astronautics. We would also like to thank the Department of Aeronautics and Astronautics at Stanford University for providing the debate on human spaceflight that motivated this paper.